\documentclass[aps,twocolumn,prl,showpacs]{revtex4-1}

\usepackage{amsfonts}
\usepackage{amsmath}
\usepackage{amssymb}
\usepackage{graphicx}

\begin{document}

\title{Spin contribution to the ponderomotive force in a plasma}

\author{G. Brodin} 
\affiliation{Department of Physics, Ume{\aa } University, SE--901 87 Ume{\aa }, Sweden}

\author{A. P. Misra} 
\affiliation{Department of Physics, Ume{\aa } University, SE--901 87 Ume{\aa }, Sweden}

\author{M. Marklund}
\email{mattias.marklund@physics.umu.se}
\affiliation{Department of Physics, Ume{\aa } University, SE--901 87 Ume{\aa }, Sweden}

\begin{abstract}
The concept of a ponderomotive force due to the intrinsic spin of electrons
is developed. An expression containing both the classical as well as the
spin-induced ponderomotive force is derived. The results are used to
demonstrate that an electromagnetic pulse can induce a spin-polarized
plasma. Furthermore, it is shown that for certain parameters, the nonlinear
back-reaction on the electromagnetic pulse from the spin magnetization
current can be larger than that from the classical free current. Suitable
parameter values for a direct test of this effect are presented.
\end{abstract}

\pacs{52.35.Mw, 52.27.-h, 52.38.-r}
\date{24 March, 2010}
\revised{17 May, 2010}
\revised{27 July, 2010}
\maketitle

The use of the spin properties of material constituents for e.g., carrying
information is currently an important paradigm \cite{Spintronics}. However,
the spin properties of the material constituents also make its presence felt
through collective effects. In particular, recent findings point to the
possibility of observing quantum plasma effects \cite{Quantum plasmas}
through the electron spin \cite{Spin-ref} in regimes otherwise thought to be
classical \cite{Classical-quant}. Such results are due to the complex
interplay between collective plasma effects and the system nonlinearities.
In classical plasmas, nonlinear effects play an important, sometimes a
crucial, role. For example, the density fluctuations induced by the
ponderomotive force of an electromagnetic (EM) wave lead to an electrostatic
wake field \cite{Wake-early}, as used in advanced particle accelerator
schemes \cite{Adv. accelerator}. In other regimes, the back-reaction on the
EM-wave due to the density fluctuations leads to phenomena such as soliton
formation, self-focusing or wave collapse \cite{NL-effects}. Such radiation
pressure-like effects are widely used in high-intensity laser experiments 
\cite{Nonlinear evolution}, and generalizations to include certain types of
quantum plasma effects have recently been made \cite{Quant-nonlin}. However,
to our knowledge the possibility of spin induced contribution to the
ponderomotive forces has not been explored.

In the present work we will solve the full set of equations for the spin
dynamics of charged particles in the presence of a weakly nonlinear EM wave
pulse, propagating parallel to an external magnetic field, in order to find
the contribution to the ponderomotive force. In the classical limit, we
recover the well-known expression first derived by Karpman and Washimi \cite%
{Karpman-pond}. The spin contribution to the ponderomotive force will in
general act in opposite directions for spin-up and spin-down populations. As
a consequence, an EM-pulse (due to, e.g., a laser or a microwave source) may
induce a spin-polarized plasma. In particular, it is demonstrated that this
mechanism can induce large spin-polarization for a laser source in the
UV-regime. For this case it should be noted that the effect of the external
magnetic field is negligible as the laser frequency is much higher than the
cyclotron frequency, but our general expression applies also for
low-frequency (lf) waves in magnetized plasmas.

When combined with the high-frequency (hf) oscillations, the classical lf
density response generates a hf current that results in a cubical
nonlinearity. This classical nonlinearity can be compared to a cubically
nonlinear magnetization current, caused by the electrons spins in a plasma
that are spin-polarized due to ponderomotive effects. It turns out that for
a plasma frequency corresponding to a metal density and a hf source (e.g.,
an x-ray free electron laser (XFEL)), the spin contribution can be larger
than the classical contribution.

We will assume the existence of two electron populations, namely spin-up and
spin-down relative to a background magnetic field $\mathbf{B}_{0}\equiv B_{0}%
\hat{\mathbf{z}}$, to be denoted by $u$ and $d$ respectively, and formally
treated as different species. Such division are relatively common in e.g.
semiconductor physics \cite{Semiconductor}. Although the spin states of the
particles will be perturbed by the presence of electromagnetic waves, the
separation of species is still well-defined provided the physics associated
with spin-flips can be neglected (see Refs. \cite{Classical-quant} and \cite%
{Spin-flip-calc} for a discussion and a calculation of spin-flip
frequencies, respectively). 
The basic equations take the form \cite{Spin-ref,Classical-quant} 
\begin{eqnarray}
&&\partial _{t}n_{\alpha }+\nabla \cdot (n_{\alpha }\mathbf{v}_{\alpha })=0,
\label{e1} \\
&&m\left( \partial _{t}+\mathbf{v}_{\alpha }\cdot \nabla \right) \mathbf{v}%
_{\alpha }=q\left( \mathbf{E}+\mathbf{v}_{\alpha }\times \mathbf{B}\right) -{%
\nabla P_{\alpha }}/n_{\alpha }  \notag \\
&&\qquad \qquad \qquad \qquad +(2\mu /\hbar )S_{\alpha }^{a}\nabla B_{a},
\label{e2} \\
&&\left( \partial _{t}+\mathbf{v}_{\alpha }\cdot \nabla \right) \mathbf{S}%
_{\alpha }=-({2\mu }/{\hbar })\mathbf{B}\times \mathbf{S}_{\alpha },
\label{e3}
\end{eqnarray}%
and $\nabla \cdot \mathbf{E}=({q}/{\varepsilon _{0}})\left(
n_{u}+n_{d}-n_{i}\right) $, where $\mathbf{S}_{\alpha }$ is the spin of
species $\alpha $ (with $\alpha =u,d$), $n_{i}$ is the ion density, and $%
q=-e $ is the electron charge, $\mu \equiv -g\mu _{B}/2$, where $\mu
_{B}\equiv e\hbar /2m$ is the Bohr magneton, $g\approx 2.0023192$ is the
electron $g$-factor, and $P_{\alpha }$ is the pressure \cite{footnote1}. The
principal condition for the validity of Eqs. (\ref{e1})-(\ref{e3}) is that
characteristic spatial scale lengths is longer than the thermal de Broglie
length. Naturally Eqs. (\ref{e1})-(\ref{e3}) also has the inherent
limitations of a fluid model, i.e. resonance effects for particle velocities
matching the phase velocity and/or group velocity are left out. Furthermore,
effects due to the off-diagonal componenets of the pressure tensor are
omitted. In what follows we consider an electron--proton plasma with
classical ion dynamics. 

Our derivation of the ponderomotive force below will be based on a direct
perturbative analysis of Eqs.\ (\ref{e1})-(\ref{e3}). We note that an
alternative approach based on Lagrangian method \cite%
{Bauer-1995,Lagrange-magn} can be useful. Assuming a slowly varying plane EM
wave $\mathbf{E=}\widetilde{\mathbf{E}}\exp [i(kz-\omega t)]+\mathrm{c.c.}$,
where c.c.\ denotes complex conjugate, 
Eq.\ (\ref{e2}) becomes 
\begin{equation}
\left( \partial _{t}-i\omega \right) \mathbf{v}_{\alpha }=({q}/{m})\mathbf{E}%
-\Omega \mathbf{v}_{\alpha }\times \mathbf{\hat{z},}  \label{e5}
\end{equation}%
where $\Omega \equiv eB_{0}/m$ is the electron-cyclotron frequency. Here we
have assumed that the background state contains no net drifts. Note that
kinetic effects, which is of potential importance (see e.g. \cite%
{Kinetic-refs}) but is outside the present model, is assumed to be
negligible throughout the manuscript. Within a fluid model, the average
longitudinal velocity, that may contribute through the convective derivative
(cf. Ref. \cite{Bauer-1995}) becomes second order in the amplitude, and
hence can be neglected here. For notational convenience, we have dropped the
tilde denoting the envelope function, and it is understood that all
derivatives act on the slowly varying amplitudes. Next, we define the
variables $v_{\alpha \pm }\equiv v_{\alpha x}\pm iv_{\alpha y}$, $E_{\pm
}\equiv E_{x}\pm iE_{y}$. 
Substituting the lowest order result $v_{\alpha \pm }={iqE_{\pm }}/{m\left(
\omega \pm \Omega \right) }$ into the correction term in Eq.\ (\ref{e5}), 
we obtain 
\begin{equation}
v_{\pm }\equiv v_{\alpha \pm }=\frac{q}{m}\frac{1}{\left( \omega \pm \Omega
\right) }\left[ iE_{\pm }+\frac{1}{\omega \pm \Omega }\frac{\partial E_{\pm }%
}{\partial t}\right] ,  \label{e7}
\end{equation}%
Using Faraday's law, $\nabla \times \mathbf{E}=-\partial_t \mathbf{B}$, we
similarly obtain the expression for the perturbed magnetic field as 
\begin{equation}
B_{\pm }=\pm \frac{ik}{\omega }E_{\pm }\pm \frac{1}{\omega }\frac{\partial
E_{\pm }}{\partial z}\pm \frac{k}{\omega ^{2}}\frac{\partial E_{\pm }}{%
\partial t}.  \label{e8}
\end{equation}%
The classical ponderomotive force component is 
\begin{equation}
F_{\mathrm{c}z} \equiv \frac{q}{m}\langle \mathbf{v}\times \mathbf{B}\rangle
_{z} =\left\{ 
\begin{array}{c}
\displaystyle{\frac{iq}{2m}\left( v_{+}B_{+}^{\ast }-v_{+}^{\ast
}B_{+}\right) }\text{ for RCP,} \\[2mm] 
\displaystyle{\frac{iq}{2m}\left( v_{-}^{\ast }B_{-}-v_{-}B_{-}^{\ast
}\right) }\text{ \ for LCP}.%
\end{array}%
\right.  \label{e9}
\end{equation}%
%
%
%
%
%
%
%
%
Substitution of Eqs.\ (\ref{e7}) and (\ref{e8}) into Eq. (\ref{e9}) gives 
\begin{equation}
F_{\mathrm{c}z}=-\frac{e^{2}}{2m^{2}\omega \left( \omega \pm \Omega \right) }%
\left[ \frac{\partial }{\partial z}\pm \frac{k\Omega }{\omega \left( \omega
\pm \Omega \right) }\frac{\partial }{\partial t}\right] |E|^{2}.  \label{e10}
\end{equation}%
in agreement with the classical result \cite{Karpman-pond}.

Next, we derive the effects due to the finite magnetic moment of the
electrons. Through the force $F_{\alpha z}\equiv ({2\mu }/{m\hbar })\langle
S_{\alpha }^{a}\nabla B_{a}\rangle _{z}$ in the averaged momentum equation,
a ponderomotive effect due to spin will be generated, where the \textit{EM
wave} will lead to the separation of the spin-up and down electrons, as will
be shown below. Starting from the linearized spin-evolution equation 
\begin{equation}
\left( \partial _{t}-i\omega \right) \mathbf{S}_{\alpha }=-({2\mu }/{\hbar }%
)\left( B_{0}\mathbf{\hat{z}}\times \mathbf{S}_{\alpha }+S_{0\alpha }\mathbf{%
B}\times \mathbf{\hat{z}}\right) ,  \label{e11}
\end{equation}%
where $S_{0u}=\hbar /2=-S_{0d}$, the contribution from the magnetic dipole
force can be obtained. Neglecting the slow time derivative, \ Eq. (\ref{e11}%
) gives 
\begin{equation}
S_{\alpha \pm }\equiv S_{\alpha x}\pm iS_{\alpha y}=\mp {2\mu S_{0\alpha
}B_{\pm }}/[{\hbar \left( \omega \pm \omega _{g}\right) }],  \label{e12}
\end{equation}%
where $\omega _{g}\equiv g\mu _{B}B_{0}/\hbar =(g/2)\Omega $ is the
spin-precession frequency. Then, including the first order correction, the
expression for the perturbed spin becomes 
\begin{equation}
S_{\alpha \pm }=\frac{2\mu S_{0\alpha }}{\hbar \left( \omega \pm \omega
_{g}\right) }\left[ \mp B_{\pm }\pm \frac{i}{\left( \omega \pm \omega
_{g}\right) }\frac{\partial B_{\pm }}{\partial t}\right] .  \label{e13}
\end{equation}%
The spin-ponderomotive force can be written as 
\begin{equation}
F_{\alpha z}=(2\mu /m\hbar )(S_{\alpha \pm }\partial _{z}B_{\pm }^{\ast
}+S_{\alpha \pm }^{\ast }\partial _{z}B_{\pm }).  \label{e14}
\end{equation}%
Substitution of Eq. (\ref{e13}) into Eq. (\ref{e14}) gives 
\begin{equation}
F_{\alpha z}=\mp \frac{4\mu ^{2}}{m\hbar ^{2}}\frac{S_{0\alpha }}{\left(
\omega \pm \omega _{g}\right) }\left[ \frac{\partial }{\partial z}-\frac{k}{%
\left( \omega \pm \omega _{g}\right) }\frac{\partial }{\partial t}\right]
|B|^{2}.  \label{15}
\end{equation}%
The above expression applies to arbitrary EM wave propagation parallel to $%
\mathbf{B}_{0}$. 
The overall structure of the spin ponderomotive part of the force (\ref{15})
is similar to the classical part (\ref{e10}). However, there are
differences. Firstly, the frequency resonances occur at the spin precession
frequency $\omega _{g}=(g/2)\Omega \approx 1.00116\Omega $. Secondly, the
dependence on the unperturbed spin state means that spin-up and spin-down
populations are pushed in opposite directions by the spin force. Thirdly,
for frequencies well below the cyclotron frequency typically the part of the
spin contribution proportional to the time-derivative is negligible, whereas
it is crucial for the classical contribution. For frequencies well above the
cyclotron frequency, the force ratio scaling is $\left\vert F_{\alpha
z}\right\vert /\left\vert F_{\mathrm{c}z}\right\vert \equiv $ $\hbar
k(1+v_{g}/v_{p})/mv_{p}\sim $ $\hbar k/mc$ for $v_{g},v_{p}\sim c,$ where $%
v_{g(p)}$ is the group (phase) speed of the wave. 
As we will see below, even a rather weak spin-ponderomotive force,
corresponding to moderately high frequencies, can lead to large
modifications of the nonlinear dynamics in an unmagnetized plasma.

We now use the expressions for the ponderomotive forces as source terms for
longitudinal lf perturbations. We define $N_{1,2}=n_{u}\pm n_{d}$ and $%
V_{1,2}=\left( v_{u}\pm v_{d}\right) /2$. In what follows, we will also
neglect any difference in the unperturbed spin populations, i.e., we will
use $n_{0u}=n_{0d}\equiv n_{0}/2$, which is a good approximation when the
Zeeman energy is smaller than the thermal energy. From the lf parts of the
continuity equations for spin-up ($u$) and spin-down ($d$) populations we
then have 
\begin{equation}
\partial _{t}N_{1,2}=-n_{0}\partial _{z}V_{1,2},  \label{e15}
\end{equation}%
From the momentum balance equations, using Eq. (\ref{e8}), we obtain for the
lf response the equations 
\begin{equation*}
\frac{\partial V_{1}}{\partial t}=\frac{q}{m}E_{l}-\frac{q^{2}}{2m^{2}\omega
\left( \omega \pm \Omega \right) }\left[ \frac{\partial }{\partial z}\pm 
\frac{k\Omega }{\omega \left( \omega \pm \Omega \right) }\frac{\partial }{%
\partial t}\right] |E|^{2},
\end{equation*}%
%
and 
\begin{equation}
\frac{\partial V_{2}}{\partial t}=\mp \frac{4\mu ^{2}k^{2}S_{0}}{m\hbar
^{2}\omega ^{2}\left( \omega \pm \omega _{g}\right) }\left[ \frac{\partial }{%
\partial z}-\frac{k}{\left( \omega \pm \omega _{g}\right) }\frac{\partial }{%
\partial t}\right] |E|^{2},  \label{e17}
\end{equation}%
where $E_{l}$ is the lf part of the electric field, $S_{0}=\hbar /2$ and $%
\omega _{p}=({n_{0}q^{2}}/{m\varepsilon _{0}})^{1/2}$. Here we have
neglected thermal effects, which is justified if $v_{\mathrm{th}}^{2}\ll
v_{g}^{2}$, where $v_{\mathrm{th}}$ is the thermal velocity. Similarly
particle dispersive effects and Fermi pressure effects, which may influence
the longitudinal dynamics \cite{Wake-field}, has been neglected, which can
be justified in the example considered below \cite{Longitudinal-note}. With
immobile positive charge carriers, Poisson's equation is $\partial
_{z}E_{l}=(q/\varepsilon _{0})(N_{1}-n_{0})$. Together with Eqs.\ (\ref{e15}%
)--(\ref{e17}), we then obtain the wave equations 
\begin{equation}
v_{g}^{2}\frac{\partial ^{2}N_{1}}{\partial \xi ^{2}}+\omega _{p}^{2}N_{1}=%
\frac{\varepsilon _{0}\omega _{p}^{2}}{2m\omega \left( \omega \pm \Omega
\right) }\left[ 1\mp \frac{kv_{g}\Omega }{\omega \left( \omega \pm \Omega
\right) }\right] \frac{\partial ^{2}|E|^{2}}{\partial \xi ^{2}},
\label{e19b}
\end{equation}%
%
%
%
%
%
%
%
%
\begin{equation}
v_{g}^{2}\frac{\partial ^{2}N_{2}}{\partial \xi ^{2}}=\pm \frac{\varepsilon
_{0}\omega _{p}^{2}k^{2}S_{0}}{2m^{2}\omega ^{2}\left( \omega \pm \omega
_{g}\right) }\left[ 1+\frac{kv_{g}}{\left( \omega \pm \omega _{g}\right) }%
\right] \frac{\partial ^{2}|E|^{2}}{\partial \xi ^{2}},  \label{e20}
\end{equation}%
where we have transformed to a comoving frame, with $\xi =z-v_{g}t$. The
spin polarization [see Eq.\ (\ref{e20})], can be integrated directly to give 
$N_{2}\propto |E|^{2}$, whereas $N_{1}$ is non-locally related to $|E|^{2}$
through (\ref{e19b}), due to the possible excitation of a plasma oscillation
wakefield with a characteristic wavelength $\lambda _{p}\equiv v_{g}/\omega
_{p}$. To demonstrate that the spin effects can be significant also when $%
B_{0}\rightarrow 0$, we compare the amplitude of the total density
perturbation $N_{1}$ with the degree of spin-polarization $N_{2}$ in an
unmagnetized plasma $\omega _{g},$ $\Omega \rightarrow 0$. Furthermore, to
be specific, we consider hf \ EM waves with $v_{g},v_{p}\sim c$. Finally, we
use $\omega _{p}\lesssim kc$ and use the estimate $\partial
^{2}|E|^{2}/\partial \xi ^{2}\sim $$|E|^{2}/L_{p}^{2}$, where $L_{p}\gg
k^{-1}$ is the length of the hf pulse. The degree of spin-polarization is
then $N_{2}/N_{1}\sim \hbar \omega _{p}(kL_{p})^{2}/mc^{2}$. We note that
the omission of ion density dynamics constrains this expression to pulse
lengths fulfilling $L_{p}\lesssim c/\omega _{pi}$, where $\omega _{pi}$ is
the ion plasma frequency. Thus, here we will consider the case of an
EM-pulse interacting with a plasma without positive mobile charge carriers,
i.e., a metal with $\omega _{p}/2\pi \simeq 10^{16}\,\mathrm{s}^{-1}$. A
numerical example with a UV-laser of wavelength, $\lambda =80\,\mathrm{nm}$
and pulse length, $L_{p}=15\,\mathrm{\mu m}$ leading to moderate
spin-polarization ($N_{2}/N_{1}\approx 3$ at the centre) is displayed in
Fig.\ 1. A longer pulse length or a shorter wavelength will give a higher
degree of spin-polarization, i.e. a strongly spin-polarized plasma with $%
N_{2}\gg N_{1}$ can be reached. We note that the polarization of the EM wave
is crucial. In the limit considered here ($\omega \gg \Omega $), the spin
contribution to the ponderomotive force has opposite direction for RCP and
LCP waves. Thus, an experiment along these lines must use circular polarized
rather than linearly polarized light, as the effect is a factor $\Omega
/\omega $ smaller in the latter case. A related use of the ponderomotive
force for isotope separation has been suggested in Ref.\ \cite%
{Weibel-isotope}, where the different charge-to-mass ratios of different
isotopes was used.

Next, we compare the back-reaction on the EM-pulse, induced by the classical
density perturbation $N_{1}$ and its spin-polarized counterpart $N_{2}$. 
From the classical current $\mathbf{J}=q(n_{u}\mathbf{v}_{u}+n_{d}\mathbf{v}%
_{d})=qN_{1}\mathbf{v}$, we have 
\begin{equation}
J_{\pm }=qN_{1}v_{\pm }={iq^{2}N_{1}E_{\pm }}/[{m\left( \omega \pm \Omega
\right) }],  \label{e21}
\end{equation}%
with $\mathbf{v}_{u}=\mathbf{v}_{d}$ for RCP and LCP waves. The
magnetization current $\mathbf{J}_{\mathrm{M}}=\nabla \times (\mathbf{M}_{u}+%
\mathbf{M}_{d})=\left( \mu /\hbar \right) \left[ \nabla \times \left( n_{u}%
\mathbf{S}_{u}+n_{d}\mathbf{S}_{d}\right) \right] $, i.e., $\mathbf{J}_{%
\mathrm{M}\pm }=\pm \left( kg\mu _{B}/2\hbar \right) \left( n_{u}\mathbf{S}%
_{u\pm }+n_{d}\mathbf{S}_{d\pm }\right) $, together with Eq.\ (\ref{e12})
and the lowest order expression of $B_{\pm }$ [see Eq.\ (\ref{e8})], gives 
\begin{equation}
J_{\mathrm{M}\pm }=\pm 8i\left( {k\mu }/{\omega \hbar }\right) ^{2}\left[ {%
\omega S_{0}}/({\omega \pm \omega _{g}})\right] \left( N_{2}E_{\pm }\right) .
\label{e27}
\end{equation}%
Now, Eq. (\ref{e20}) can be integrated for $N_{2}$ . A Gaussian pulse $%
|E|=E_{0}\exp (-\xi ^{2}/L_{p}^{2})$ gives (for $B_{0},\omega
_{g}\rightarrow 0$) 
\begin{equation}
N_{2}=\pm \frac{\varepsilon _{0}\omega _{p}^{2}k^{2}S_{0}|E_{0}|^{2}\exp
(-2\xi ^{2}/L_{p}^{2})}{m^{2}\omega ^{3}v_{g}^{2}}\left( 1+\frac{v_{g}}{v_{p}%
}\right) .  \label{e30}
\end{equation}%
Furthermore, for a pulse length much larger than the plasma oscillation
wavelength $\lambda _{p},$ we can use the following estimate from Eq. (\ref%
{e19b}) as $(B_{0},\Omega \rightarrow 0)$ 
\begin{equation}
N_{1}\sim \lbrack {\varepsilon _{0}\omega _{p}^{2}|E_{0}|^{2}\exp (-2\xi
^{2}/L_{p}^{2})}]/({m\omega ^{2}v_{g}^{2}L_{p}^{2}k_{p}^{2}}).  \label{e31}
\end{equation}%
The density ratio for RCP and LCP waves is then given by 
\begin{equation}
\left\vert \frac{N_{2}}{N_{1}}\right\vert \sim \left( \frac{\hbar \omega _{p}%
}{mc^{2}}\right) \left( kL_{p}\right) ^{2}\left( \frac{c}{v_{p}}\right)
^{2}\left( \frac{\omega \omega _{p}}{k^{2}v_{g}^{2}}\right) \left( 1+\frac{%
v_{g}}{v_{p}}\right) ,  \label{e32}
\end{equation}%
where $k_{p}\equiv 1/\lambda _{p}.$ The ratio of the two currents for RCP
and LCP waves is then given by $\Gamma \equiv |J_{\mathrm{M}\pm }/J_{\pm
}|\approx ({\hbar \omega }/{mv_{p}^{2}})|N_{2}/N_{1}|$, i.e. 
\begin{equation}
\Gamma \sim \left( {\hbar \omega _{p}}/{mc^{2}}\right) ^{2}\left(
kL_{p}\right) ^{2}\left( {c^{2}}/{v_{p}v_{g}}\right) ^{2}\left( 1+{v_{g}}/{%
v_{p}}\right) .  \label{e33}
\end{equation}%
Thus, for $v_{g},v_{p}\sim c$, we have 
\begin{equation}
\Gamma \sim \left( {\hbar \omega _{p}}/{mc^{2}}\right) ^{2}\left(
kL_{p}\right) ^{2}.  \label{e34}
\end{equation}%
Solving Eqs.\ (\ref{e19b}) and (\ref{e20}), we in Fig.\ 2 compare the two
current profiles for an XFEL with $\lambda =1\,\mathrm{nm}$, a pulse length $%
L_{p}=30\,\mathrm{\mu m}$, and a metallic plasma density, giving $\omega
_{p}/2\pi =10^{16}$ $\mathrm{s}^{-1}$. Our estimate (\ref{e34}) is then
verified, and it is found that the central value of the current ratio is $%
\Gamma \approx 3$. These parameters are relevant for the XFEL at DESY \cite%
{XFEL-DESY}. In fact, the shortest wavelength generated by this facility is $%
\lambda =0.1$\textrm{nm}, making the quantum mechanical back-reaction much
larger than the classical response [$\Gamma \sim 200$ according to Eq. (\ref%
{e34})].

\begin{figure}[tbp]
\includegraphics[width=0.75\columnwidth]{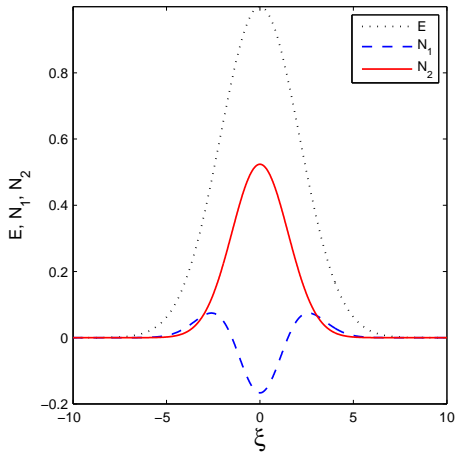}
\caption{The normalized classical density perturbation $N_{1} $ and the
spin-induced density difference $N_{2}$ together with a Gaussian EM-pulse, $%
|E|=|E_{0}|\exp (-\protect\xi ^{2}/L_{p}^{2})$, as calculated from Eqs.\ (%
\protect\ref{e19b}) and (\protect\ref{e20}). The parameters correspond to an
unmagnetized plasma with $\protect\omega_p/2\protect\pi = 10^{16}\, \mathrm{s%
}^{-1}$, $\protect\lambda =80 \, \mathrm{nm}$, and $L_{p}=15 \, \mathrm{%
\protect\mu m}$ (a.u.).}
\end{figure}

\begin{figure}[tbp]
\includegraphics[width=0.85\columnwidth]{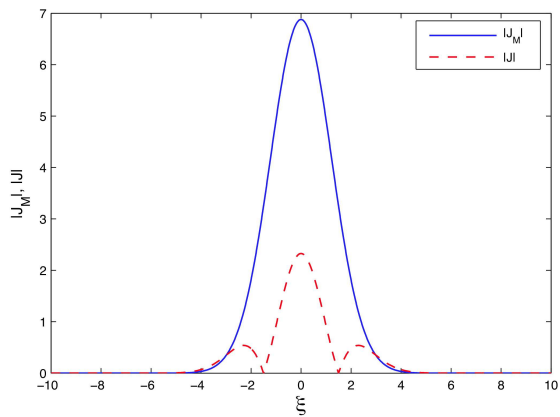}
\caption{The profiles of the normalized spin-induced current density $J_{M}$
and the classical current density $J$ for a Gaussian EM-pulse, as obtained
from Eqs. (\protect\ref{e21}) and (\protect\ref{e27}). The parameters
correspond to an unmagnetized plasma with $\protect\omega_p/2\protect\pi =
10^{16}\, \mathrm{s}^{-1}$, $\protect\lambda =1 \, \mathrm{nm}$, and $%
L_{p}=30 \, \mathrm{\protect\mu m}$ (a.u.).}
\end{figure}


In the present Letter, we have generalized the classical expression for the
ponderomotive force in a magnetized plasma to include the effect of the
electron spin. Our main result, Eq. (\ref{15}), applies for arbitrary
electromagnetic waves propagating along an external magnetic field. One of
the main features of the spin-ponderomotive force is that it can induce a
strong spin-polarization in a plasma, even if the initial up- and down-
states of electrons are equally populated. An example with an EM-pulse in
the UV-regime is given in Fig.\ 1. Furthermore, even in an unmagnetized
plasma, the nonlinear back-reaction from the spin-induced current can be
larger than the classical back-reaction, provided the EM-pulse has a
sufficiently short-wavelength. It should be stressed that we have here only
compared the spin-polarized current response with the classical current
density, and that relativistic nonlinearities (see e.g. \cite{NL-effects})
that has been omitted here may play a role for the full dynamic evolution.
An example with an XFEL is given in Fig.\ 2. Finally, we want to point out
that the possibilities of nonlinear spin effects is still a relatively
unexplored area, and generalizations, such as arbitrary directions of
propagation, are likely to lead to new and interesting discoveries.

A.P.M.\ acknowledges support from the Kempe Foundations and thanks
B.\ Eliasson for help regarding numerical simulations. This research
is supported by the European Research Council and the Swedish Research Council under Contracts \#
204059-QPQV and 2007-4422.




\begin{thebibliography}{99}
\bibitem{Longitudinal-note} The Fermi pressure and the Bohm-de Broglie
potential has been shown to influence the longitudinal dynamics excited by
the ponderomotive force \cite{Wake-field}. However, for the case when $%
v_{g}\gg v_{F}$ (where $v_{F}$ is the Fermi velocity), the physics on length
scales much longer than the Compton wavelength, which here is our concern,
is unaffected.
\end{thebibliography}
\end{document}